# A Heterogeneous General Model for Neuromorphic-Inspired Computation

M. Mirigliano[1]

[1]*Satenlight, Via Timavo 34, Milano*

**Abstract**: *In recent years, both academia and industry have focused on the development of computational architectures inspired by the distributed, adaptive, and event-driven characteristics of biological neural systems, with the aim of reducing the computational cost associated with conventional training approaches* [1]. *However, a major challenge is the lack of general models and design guidelines for emerging computational systems and hardware. This work introduces a general model based on an input-dependent stochastic weight network, referred to as a substrate. The substrate weights evolve through input-triggered stochastic updates, with correlations between weight coefficients described by a matrix-valued covariance kernel. The proposed framework is implemented using quadratic polynomial weight functions, where the input amplitude controls the magnitude of the stochastic perturbation and a substrate-dependent distance determines the correlation structure. Numerical simulations show that correlations in the stochastic weight evolution significantly affect the system response, suggesting a potential mechanism for neuromorphic-inspired computation without conventional weight training. The aim of this work is to provide a general formulation of the model and identify its main properties and characteristics.*

## 1. Introduction

Over the last decades, data collection and processing have pervaded virtually every field of human activity, from environmental monitoring [2] and security [3] to industrial automation [4], financial activities [5], and even recreational applications such as streaming [6]. As a result, modern societies have become strongly dependent on data-processing technologies. This has been made possible, on one hand, by the high throughput and scalability achieved by silicon-based hardware, which has enabled the development of high-performance systems based on AI models [1]. On the other hand, in recent years this paradigm has revealed significant limitations, particularly in terms of energy efficiency [7] and along the entire production chain (design, fabrication, and integration) [8].

Biologically inspired systems, most notably the human brain, have emerged as a source of inspiration for new hardware, materials, and architectures, owing to their intrinsic efficiency in performing tasks such as image recognition and speech processing[1], [7]. This has driven research in two complementary directions: on one hand, toward the redesign and re-implementation of silicon-based technologies [9]; on the other, toward the fabrication and study of novel systems exhibiting nonlinear and complex responses to external stimuli of different nature—electrical [10], optical [11], or chemical [12], [13]-with the aim of enhancing the processing capability of external inputs for complex tasks such as image or speech recognition. Among these systems, those based on the self-assembly of sub-units such as nanowires [14], nanoparticles [15], and clusters [16], [17] represent a particularly promising alternative to standard technologies, owing to their fabrication simplicity and their compatibility with large-scale integration in industrial manufacturing processes.

The main limitations of these systems lie in the absence of a general theoretical framework to guide the design and development of new hardware, and in their intrinsically random, multi-level response to external electrical stimuli. These characteristics make it difficult to interface such systems with standard information-processing paradigms—such as those based on silicon technology—which rely on a digital representation of information and therefore impose strong constraints on the implementation of discrete '0' and '1' states in hardware [1].

In this work, I present a model inspired by the behavior of this class of materials, aimed at implementing unconventional information processing. The proposed model can be regarded as a formalization and generalization of the model presented in [16], [18]. It is based on the mapping of inputs onto a space of variable dimensionality through a nonlinear, input-dependent weight matrix. In addition, the intrinsically random nature of the system's response to external inputs above a given threshold is exploited to implement adaptive behavior capable of solving specific tasks. A software implementation illustrating the main features of the model is also presented. This work does not aim to benchmark the model's performance against other models implementing similar tasks or information-processing schemes, nor to demonstrate a hardware implementation of a physical system realizing the proposed architecture; these aspects will be addressed in future publications.

The paper is organized as follows. Section 2 presents the model in its general form. Section 3 describes a software implementation based on simplifying assumptions, illustrated through a system tracking target value with time-varying inputs. Section 4 presents the results, followed by a discussion in Section 5. Conclusions are summarized in the final section.

## 2. The Model

A schematic representation of the network model is shown in Figure 1.

*Figure 1: Schematic representation of the proposed network model. The non-linear input dependent weight matrix, called substrate, is shown mapping input $(x_1, .., x_N)$ into output $(o_1, \dots, o_M)$. The output is then elaborated by a fixed matrix called reduction layer (RL) resulting in the vector $(c_1, \dots, c_k)$. The system output R is computed linearly combining the elements of $(c_1, \dots, c_k)$ through the Linear Combinator LC.*

The main element is a non-linear input dependent weight network that maps an N-dimensional input array to M-dimensional output array performing a weighted sum of the input components. I refer through the text to this network as *substrate*, represented by the weight matrix:

$$\boldsymbol{W}(\boldsymbol{x}, \boldsymbol{A}_t) = \begin{pmatrix} w_{11,\boldsymbol{A}_{11,t}}(x_1, \dots, x_N) & \cdots & w_{1M.,\boldsymbol{A}_{1M,t}}(x_1, \dots, x_N) \\ \vdots & \ddots & \vdots \\ w_{N1,,\boldsymbol{A}_{N1,t}}(x_1, \dots, x_N) & \cdots & w_{NM,,\boldsymbol{A}_{NM,t}}(x_1, \dots, x_N, t) \end{pmatrix} \quad (1)$$

where weights $\boldsymbol{w}_{ij,A_{ij}}$ are non-linear functions of the inputs and they are parametrized by D coefficients represented by the vector $\boldsymbol{A}_{ij,t}$ and *t* represents the step of realization. Input is represented as a vector $\boldsymbol{x} = (x_1, \dots, x_N)$ of space $\mathcal{R}^N$ and output $\boldsymbol{o} = (o_1, \dots, o_M)$ by an array of the space $\mathcal{R}^M$. The mapping is realized through a weight matrix:

$$\boldsymbol{o} = \boldsymbol{x}W(\boldsymbol{x}, \boldsymbol{A}_t) \tag{2}$$

The distinctive feature of the substrate is that its coefficients are not obtained through a conventional training or optimization procedure. Instead, they evolve according to an input-triggered stochastic process in D-dimensional real space: $\boldsymbol{A}_{\boldsymbol{ij,t}} = \left\{A_{ij,t_{\boldsymbol{1}}}, \dots, A_{ij,t_{\boldsymbol{D}}}\right\}$. When one or more components of the external input exceed a prescribed threshold, the corresponding coefficient vectors undergo a stochastic jump and a new realization is sampled. Thus, the evolution of the substrate can be interpreted as a driven stochastic process, or more specifically as an input-triggered jump process [19].

Importantly, the stochastic evolution of different coefficients is not assumed to be independent. A perturbation affecting one coefficient may induce correlated changes in other coefficients of the substrate. This correlation is described by an input-dependent covariance kernel K

$$Cov(\boldsymbol{A}_{ij,t}, \boldsymbol{A}_{sq,t} | \boldsymbol{x_t}) = \boldsymbol{K}((i,j),(s,q),\boldsymbol{x}_t) \tag{3}$$

where $K\big((i,j),(s,q),\boldsymbol{x}_t\big)$ is a covariance kernel that defines the correlations across the substrate and $\boldsymbol{x}_t$ the input at temporal step t. The kernel is input dependent and can be decomposed into an amplitude term, a spatial or topological correlation term, and a covariance matrix describing the correlations between the D coefficients:

$$\boldsymbol{K}\big((i,j),(s,q),\boldsymbol{x}_t\big) = \sigma^2(\boldsymbol{x})\kappa\left(\frac{d\big((i,j),(s,q)\big)}{l(x)}\right)\Sigma \tag{4}$$

Here, $\sigma(\boldsymbol{x})$ is an input-dependent amplitude parameter controlling the magnitude of the stochastic variation; $d\big((i,j),(s,q)\big)$is a distance measure defined on coefficient indices $(i,j)$ and $(s,q)$ ; $l(x)$ is an input-dependent correlation length controlling the spatial or topological extent of the correlation, $\kappa(.)$ is a normalized scalar correlation function, and is $\sum$ a D × D positive semidefinite covariance matrix describing the statistical correlation between the parameters defining an individual weight. The parameter $\sigma(\boldsymbol{x})$ determines how strongly the input modifies the substrate. A small value of $\sigma(\boldsymbol{x})$ corresponds to small stochastic variations of the coefficients, while a large value produces larger deviations from their current values. Conversely, $l(x)$determines how broadly the input-induced perturbation propagates through the substrate. A small value of $l(x)$produces a localized modification, whereas a large value produces correlated changes extending over a larger portion of the substrate. Under this framework, no conventional deterministic training procedure is therefore required. Instead, the substrate explores its weight space through stochastic transitions whose probability distribution is determined by the input-dependent matrix-valued covariance kernel. The evolution of the substrate is consequently governed by three related quantities: $\sigma(\boldsymbol{x})$, which determines how much the weights change; $l(x)$, which determines how broadly the change propagates across the substrate; and $\sum$, which determines how the different parameters describing each weight are correlated. As an example, a Gaussian covariance kernel can be defined as: $K\big((i,j),(s,q),x_t\big) = \sigma^2(\boldsymbol{x})e^{-\frac{(i-s)^2}{l(x)}}$, if the dominant locality is along the input dimensions; as an alternative, one possible definition of distance function is $\mathrm{d}\big((\mathrm{i},\mathrm{j}),(\mathrm{s},\mathrm{q})\big) = \mathrm{d}_{in}(\mathrm{i},\mathrm{s}) + \eta\mathrm{d}_{out}(\mathrm{j},\mathrm{q})$. In the next section a practical realization under simplifying assumptions will be given.

Once the substrate output vector ***o*** is obtained, it is mapped into a vector with reduced dimensionality by a *reduction layer* ***RL:***

$$(o_1, \dots, o_M)\begin{pmatrix} rl_{11} & \cdots & rl_{1K} \\ \vdots & \ddots & \vdots \\ rl_{M1} & \cdots & rl_{MK} \end{pmatrix} = \begin{pmatrix} c_1 \\ \vdots \\ c_K \end{pmatrix} \tag{5}$$

The resulting vector ***c*** is called *control vector.* The reduction layer is a fixed coefficient matrix and no training process is required, the coefficient values are chosen a-priori, according to the task. Its main role is to reduce the dimensionality of the substrate output. This can be carried out performing mathematical operations like an average of different elements of vector ***o*** or cancelling some the elements of ***o***.

The control vector is then mapped into a scalar value by a layer called *Linear Combinator*:

$$(c_1, \dots, c_K)\begin{pmatrix} \alpha_1 \\ \vdots \\ \alpha_K \end{pmatrix} = R \tag{6}$$

Layer coefficients are chosen a-priori, and they realize a linear combination of the elements of the control vector to give a scalar result. The result R can be treated as an analogue signal or, alternatively, it can be given to a thresholding function to obtain a digital value:

$$\Theta(R) = \begin{cases} 1 & if\ R > s \\ 0 & otherwise \end{cases} \tag{7}$$

## 3. Examples of Application

This section presents a software implementation of the model under some simplifying assumptions and uses it to solve a simple problem in order to show an example application. The aim is to study the behaviour of the substrate when a practice problem is demanded to be solved and to study the effects of the main features of the network: non-linearity, random update and correlations. The example application will be a feedback-based control system that carries out weight adjustment to follow a target value (*set point* or SP) within an interval of acceptance defined by a tolerance (or *tol*) parameter with randomly evolving inputs.

The i-th element of the input vector $x_{i,t}$ at temporal step t is computed by this formula:

$$x_{i,t} = \alpha x_{i,t-1} + (1-\alpha)\mu + \varepsilon * s \tag{8}$$

where $0 \le \alpha < 1$ and determine a memory effect, $\mu$ is a random variable uniformly distributed between -5 and 5, and $\varepsilon$ is a random noise uniformly distributed between 0 and 1, $s$ is a constant used to tune noise scale. The response of the network is studied in different conditions:

- With $\alpha = 0.2$ and $\alpha = 0.8$
- With $s = 1$ and $s = 5$

$\alpha = 0.2$ gives weak temporal correlated input signal while with $\alpha = 0.8$ strongly correlated input signal (there is a memory effect: each new sample is strongly influenced by the previous one). $s = 1$ gives a noise fluctuation of 10% with respect the fluctuations of signal evolution, while $s = 5$ gives a noise fluctuation of 50% with respect the signal one. This allows to probe the effects of noise on the output of the system.

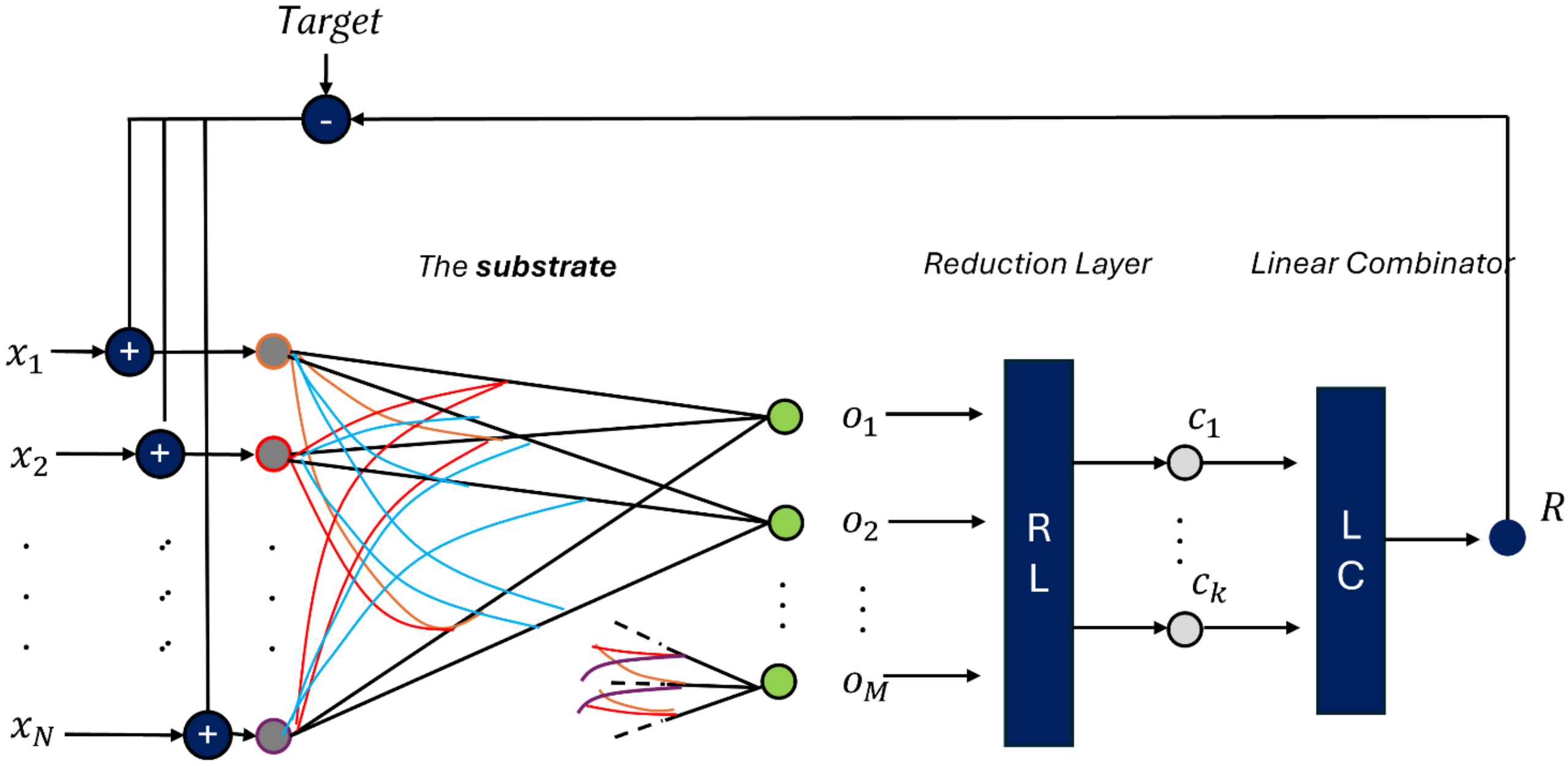

*Figure 2: The model realization in a tracking control scheme implemented through a software simulation. The output R of the system is subtracted to a Target value and the result summed to each input $x_1, \dots, x_N$.*

Figure 2 shows the schematic representation of the case of study. The example shows a substrate with N=3 inputs and M= 3 outputs. As shown in Figure 2 the difference between the desired target and output R of the system is computed and then summed to each input. In this way, when the output is far from the target, one or more inputs will overcome the threshold that trigger the weights evolution. The threshold is determined following this procedure:

- Several runs are carried out to test Input evolution following the equation (6) under different parameters, in order to study the interval explored for random evolution, for each case of study
- The maximum value of the input variation $I_{max} = \max(input) - \min(input)$ is computed
- The threshold is computed following the formula: $threhsold = SP * tol * 0.1 + I_{max}$

The threshold can also be set manually. The procedure described above is used here to make the simulation independent of the target value and to improve the generality of the results.

Now we describe in deeper detail the properties of the substrate.

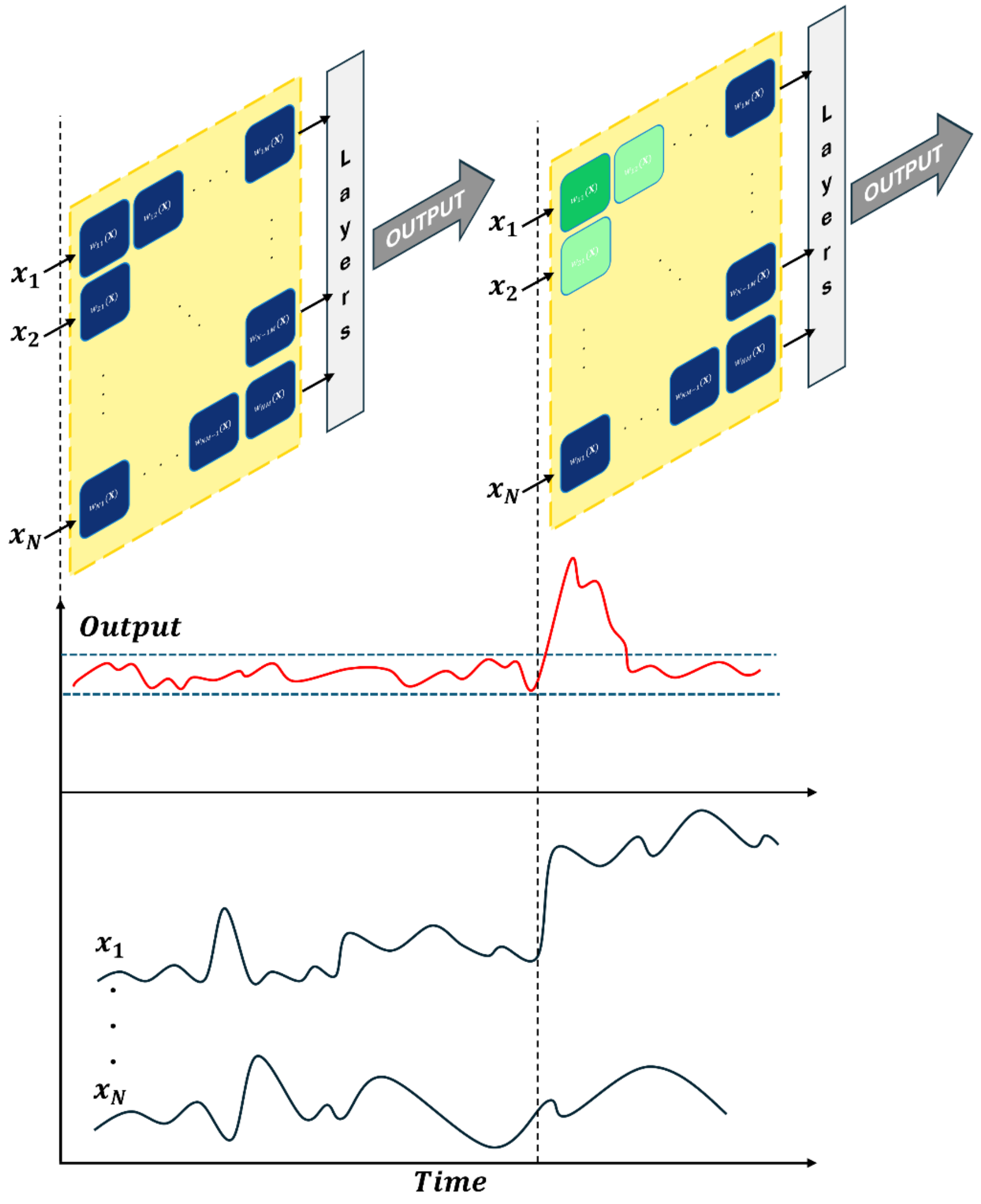

*Figure 3: Schematic representation of the switching mechanism that trigger a weight update process to follow the target value. When the output is outside a tolerance interval (manually defined in this example) one or more input will trigger the update process. The weight coefficients will be updated following a random sampling. The distance-dependent structure of the stochastic process results in larger variance for weights located near the over-threshold inputs (dark green), with the effect decreasing with distance (light green). When correlations are present, the same distance-dependent structure also induces correlated sampling among the affected weights..*

In this software implementation, the general kernel of Eq. (4) is specialized by setting $\Sigma = I_D$, where $I_D$ is the identity matrix. Thus, the *D* coefficients within each coefficient vector $\boldsymbol{A}_{ij,t}$ are assumed to be mutually uncorrelated, while correlations between different weights *(i,j)* are modelled explicitly. Accordingly, the *D* components of both the shared and idiosyncratic stochastic terms are sampled independently. Extending Σ beyond the identity matrix is left for future work

The weight functions of the substrate matrix are described as degree-2 polynomials with random coefficients:

$$w_{ij,\{\boldsymbol{A}_{ij,t}\}} = A_{ij,t}^{(0)} + \sum_{m=1}^{2} A_{ij,t}^{(m)}\, x_m + \sum_{1 \le m \le n \le N} A_{ij,t}^{(mn)} x_m x_n \tag{9}$$

where $\{\boldsymbol{A}_{ij,t}\}$ is a realization of the stochastic process with discrete time t for the weight $w_{ij}$. The number of possible coefficients for each weight $w_{ij}$ can be computed knowing the number *N* of input variables $x_1, \dots, x_N$ (corresponding to the number of polynomial variables) and the degree Z of the polynomial describing the weight functions; it can be computed by the combinatorial calculus:

$$D = \binom{N+Z}{Z} \tag{10}$$

In the present work, N= 3 inputs, polynomial degree Z = 2, it results in D=10. Polynomial of degree two are used because they are one of the simplest function forms that well capture non-linear input elaboration and input weighting process in several hardware systems [20], [21] and it is simple to implement the simulation process. The reader must be aware that the model introduce in the previous section does not limit to the polynomial case.

In this implementation, a distance measure $d_{ij,k}$ between weight *(i,j)* is defined for an input *k*:

$$d_{ij,k} = \frac{|i-k|}{2N} + \frac{|j-k|}{2M} + \epsilon \tag{11}$$

where index *k* is referred to the input and *i-j* refers to the element (or weight) of the substrate matrix, N=M and $\epsilon$ is a regularization parameter set to 0.001 to avoid the case of distance equal to 0. The defined measure expresses a sort of distance of the k-th input from the *i-j* weight of the matrix. When the k-th input overcomes a certain threshold, the substrate evolution takes place and the coefficients $\boldsymbol{A}_{ij}$ will be updated according to:

$$\boldsymbol{A}_{ij,t+1} = \rho_{ij} F_t + \sqrt{1-\rho_{ij}^2}\varepsilon_{ij} \tag{12}$$

Where the shared factor $F_t = Norm(A_{k1}, \sigma_{k1})$ drawn once per update and reused for every *(i,j)*, and the idiosyncratic term $\varepsilon_{ij} = Norm(A_{ij}, \sigma_{ij})$ drawn independently for each weight. The stochastic update contains two contributions: $\rho_{ij}$ represent the coupling between the weight *(i,j)* and the stochastic perturbation generated by the k-th input. $\sigma_{ij}$ is the variance. The distance defined in Eq. (12) is used to determine both the coupling to the shared stochastic component and the magnitude of the independent perturbation. The coupling coefficient is defined

$$\rho_{ij} = k(d_{ij}) = (1 - d_{ij,k})(-1)^{i+j} \tag{13}$$

and

$$\sigma_{ij} = \frac{1}{d_{ij,k}} * \frac{x_k}{x_{max}} \tag{14}$$

where $x_{max}$ is the max value of the input computed on several runs as described above. In this way the standard deviation is inversely proportional with respect to the distance, and it is proportional to the ratio of the k-th input and the maximum possible value of the input itself. The factor ($-1^{i+j}$) in eq. (13) is included so that both positive and negative couplings are explored. The shared term determines the covariance between different weights, whereas the independent term contributes only to the diagonal elements of the covariance matrix. Consequently:

$$Cov(A_{ij}, A_{sq}) = \rho_{ij}\rho_{sq}\sigma_{k1}^2 \ (i,j) \neq (s,q) \tag{15a}$$

$$Var(A_{ij}) = \rho_{ij}^2 \sigma_{k1}^2 - (1 - \rho_{ij}^2)\sigma_{ij}^2 \qquad (15b)$$

i.e. the $\sigma_{ij}$ term is structurally confined to the diagonal of the covariance and can never appear in an off-diagonal term, no matter how the kernel is written. Formally, this specializes in the general kernel of eq. (4) into a shared term plus an independent nugget:

$$\boldsymbol{K}((i,j),(s,q),\boldsymbol{x}_t) = \sigma_{shared}^2(\boldsymbol{x})\kappa\left(\frac{d((i,j),(s,q))}{l(x)}\right)\Sigma + \delta_{(i,j),(s,q)}\sigma_{idio}^{\mathbf{2}}\left((i,j),\boldsymbol{x_t}\right)\Sigma \qquad (4')$$

where $\delta_{(i,j),(s,q)}$ is 1 if the two indices coincide and 0 otherwise. This is a hub-based special case of the general pairwise kernel of eq. (4): correlation is mediated only through a single shared factor tied to the triggering input *k*, not through a fully general pairwise structure. Two weights that are both far from the hub but close to each other are not correlated under this construction. In the actual implementation, the general amplitude and correlation-length functions in eq. (4) are replaced by the explicit distance-dependent functions defined in Eqs. (13) and (14) In Figure 3 a schematic representation of the model is shown. The weight highlighted in dark green represent those closest to the triggering input with k=1 (lower distance) and the light green represent regions represent weights located farther away and so on for the other weights.

Note that eqs. (13) and (14) apply two different decay laws to the same distance (1−*d*) for the shared/correlation term, 1/*d* for the amplitude term — rather than a single shared *κ* governing both; eq. (4') makes this an explicit, structural feature of the model rather than an unstated inconsistency. Weights nearer the over-threshold input (input 1 in Figure 3) are therefore affected more strongly (higher $\sigma_{ij}$) and more strongly correlated with one another than weights farther away. If several inputs are simultaneously over threshold, the distance is computed with respect to each of them and all corresponding weights are updated. For comparison, simulations are also run with the coupling coefficients set identically to zero, isolating the effect of the correlations in the model.

Finally, the reduction layer is set equal to:

$$\begin{pmatrix} 1 & 0 & 0 \\ 0 & 1 & 0 \\ 0 & 0 & 1 \end{pmatrix} \qquad (16)$$

The Linear Combinator is equal to:

$$\begin{pmatrix} 1 & 1 & 1 \end{pmatrix} \qquad (17)$$

In this case the outputs of the substrate are simply summed to give the output of the system. The output does not undergo a thresholding process and it is considered as analogue.

# 4. Results

Each simulation consists of 100 time steps, with 100 input-evolution steps. In

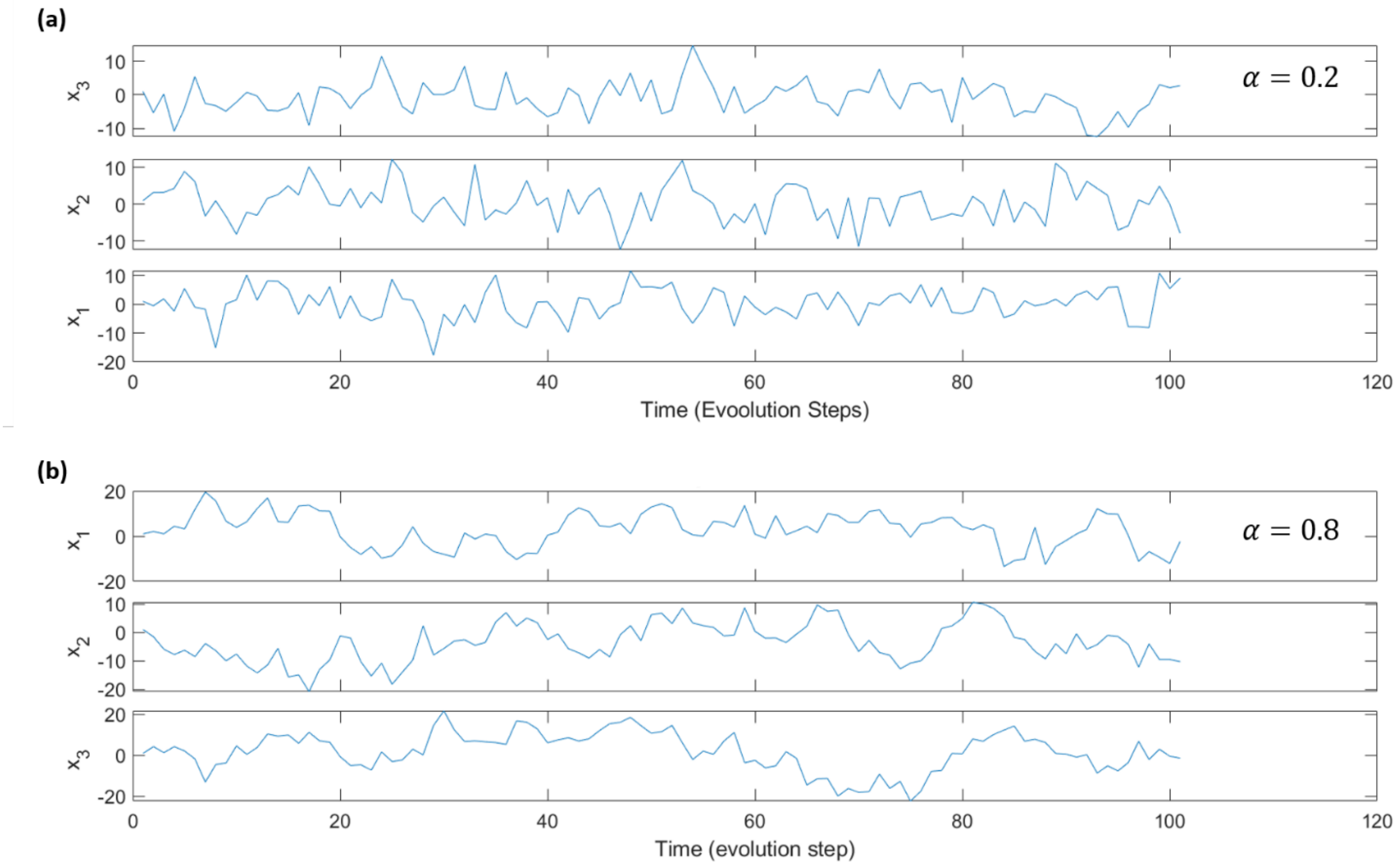

Figure 4 (a) and (b). Simulations are carried out with $\alpha = 0.2$ and $\alpha = 0.8$ and noise scale $s = 1$ and $s = 5$ (see eq. 6). Panel (a) shows the evolution with $\alpha = 0.2$ and panel-b shows the evolution with $\alpha = 0.8$. A greater parameter $\alpha$ corresponds to a longer correlation interval, in other words, the smaller the parameter $\alpha$ less correlate the input is. This can be easily seen by the autocorrelation functions that for input in panel-b shows a slower decay (graph not shown).

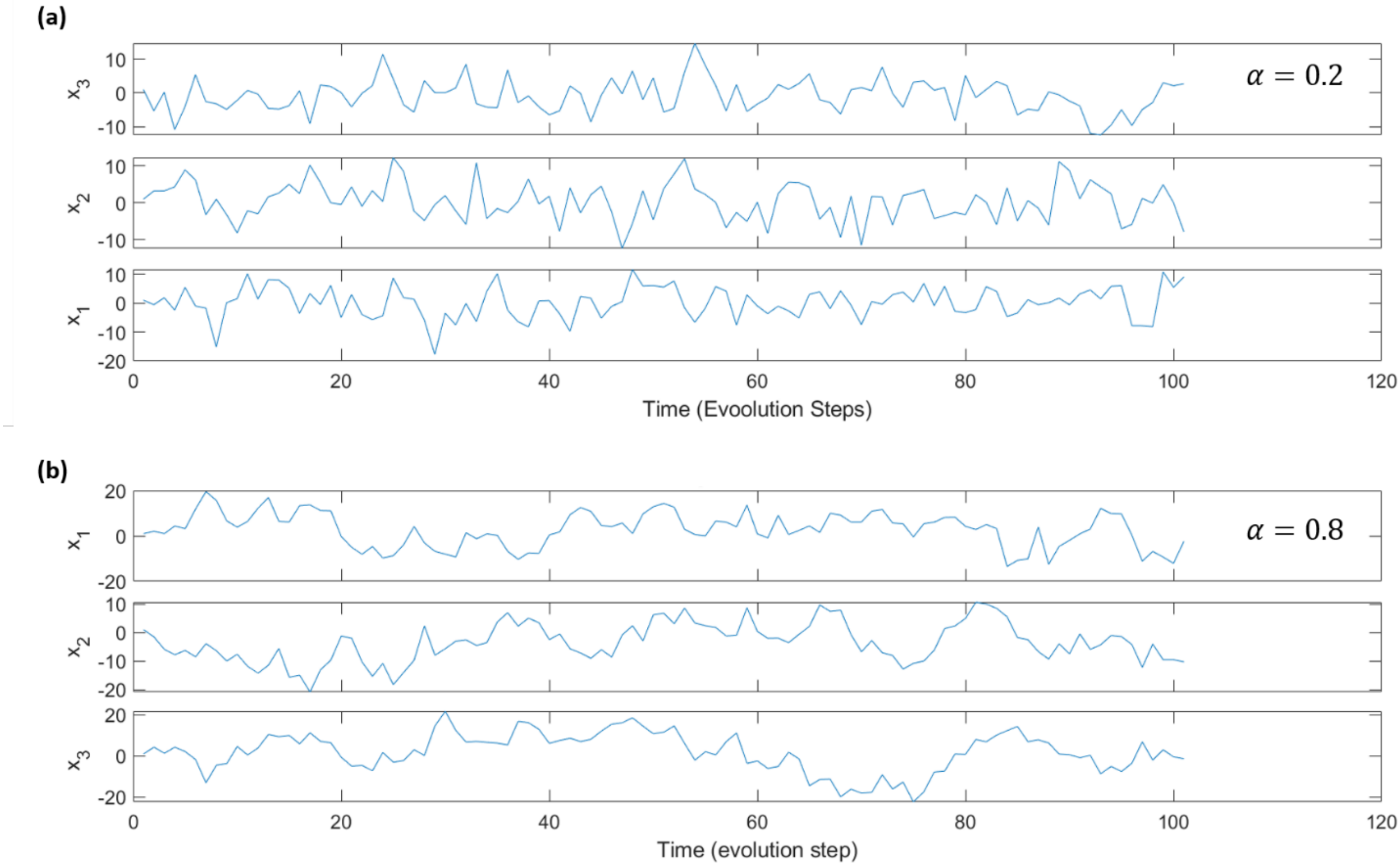

*Figure 4:example of input evolution for 100 temporal steps. (a) the input resulting tuning parameter $\alpha = 0.2$. (b) the input resulting tuning parameter $\alpha = 0.8$. At each time steps in the figure, is input is over threshold, the weight update process is started.*

At each time steps the output of the system is computed and, if the system feedback is greater than the threshold, the weight update process takes place. The update process is based on the stochastic evolution of the weight coefficients according to Eq. (13), combining a shared stochastic component associated with the triggering input and an independent component centered on the current coefficient vector of each weight. Simulations are carried out with correlations among weights as described above (see eq.12) and for the sake of comparison with correlations set to 0. The maximum number of realizations for weight update is set to 2000: if the right weight combination is not found after 2000 weight samplings, the update process stops. The update process is considered successful if at the end output of the system is in the interval $[\mathrm{SP} - \mathrm{SP} * \mathrm{tol}, \mathrm{SP} + \mathrm{SP} * \mathrm{tol}]$ . SP is set to 500 and *tol* to 10%. The weight matrix is initialized with random values that will trigger the update so in the first evolution step the weights will be updated in order that output meets SP specifications. For each simulation the number of successes (the number of update process that end with the output near to SP) and the number of realizations (the number of samplings needed to reach the output result) are computed. For each parameters combination explored (parameter $\alpha$ to tune input time correlation, parameter $s$ to set noise scale and presence of correlations in weight update process) 10 simulations have been run and the average number of successes and of realizations are computed.

| $\alpha = 0.2$ | | |
|---|---|---|
| | $s = 1$ | $s = 5$ |
| With Correlation | 90.2 | 98.8 |
| Without Correlation | 94.8 | 73.1 |

| $\alpha = 0.8$ | | |
|---|---|---|
| | $s = 1$ | $s = 5$ |
| With Correlation | 93.2 | 86.2 |
| Without Correlation | 97.5 | 47.6 |

*Table 1: Average success rate obtained for the different simulation conditions.*

| $\alpha = 0.2$ | | |
|---|---|---|
| | $s = 1$ | $s = 5$ |
| With Correlation | 276 | 202 |
| Without Correlation | 266 | 899 |

| $\alpha = 0.8$ | | |
|---|---|---|
| | $s = 1$ | $s = 5$ |
| With Correlation | 198 | 627 |
| Without Correlation | 162 | 1346 |

*Table 2: Average number of realizations obtained for the different simulation conditions*

In Table 1 the success rate for the input evolution with $\alpha = 0.2$ and $\alpha = 0.8$ are shown. IN both cases the weight update process without correlations results in a slightly higher success rate when noise scale $s$ is equal to 1. The results change dramatically with noise scale $s = 5$: in this case the evolution with correlations gives better results and in the case with $\alpha = 0.8$ the evolution without correlations fails the majority of the trials in finding the right weight combination to give the desired result within 2000 realizations. This reflects also in the number of realizations to implement the weight update: as shown in Table 2 in the case with correlations and noise scale $s = 5$ is much lower than the case without correlations. I highlight that the case with $s = 5$ (higher noise) and $\alpha = 0.8$ (higher temporal correlations in input evolution) is the hardest to be resolved by the system.

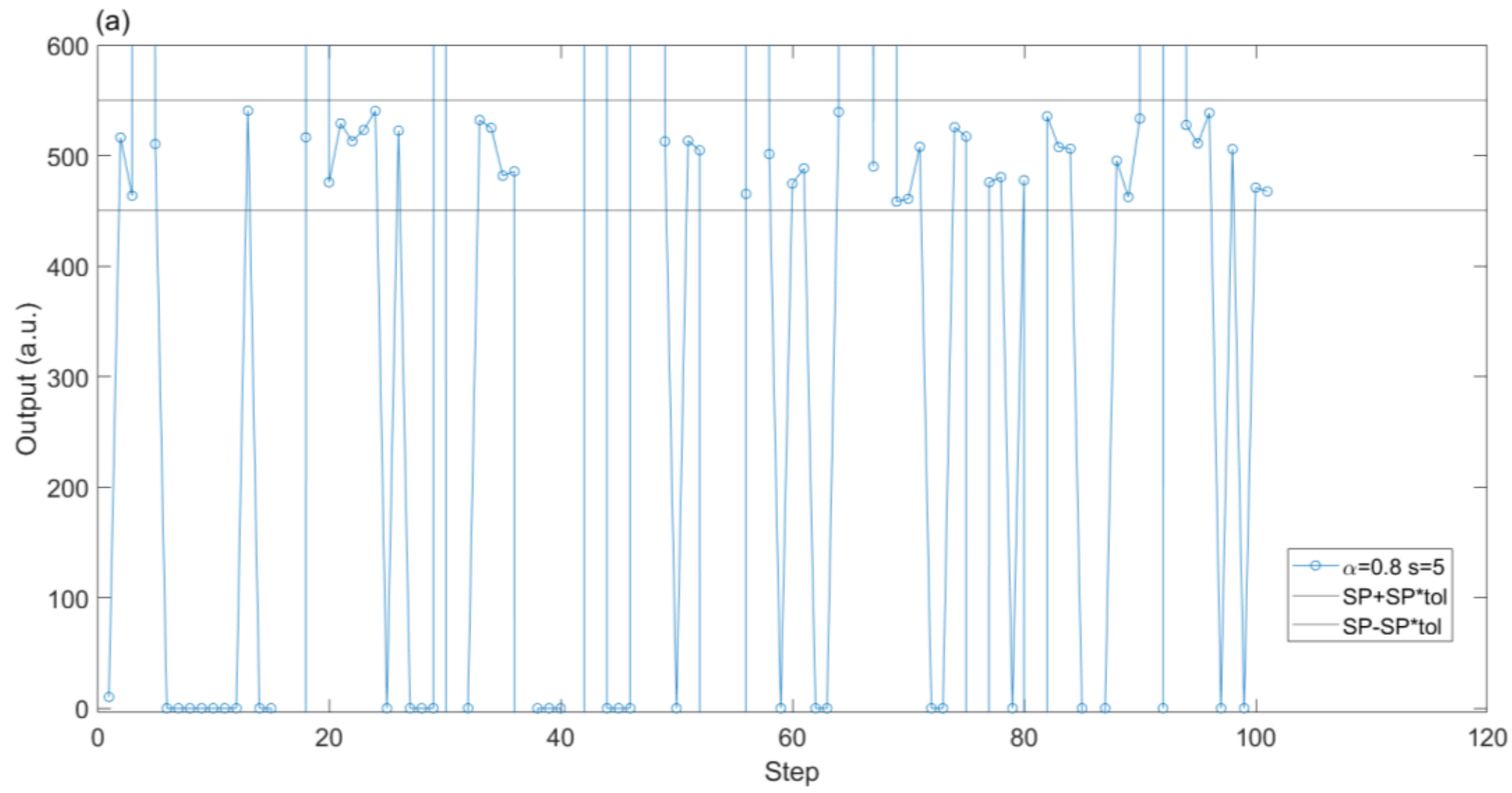

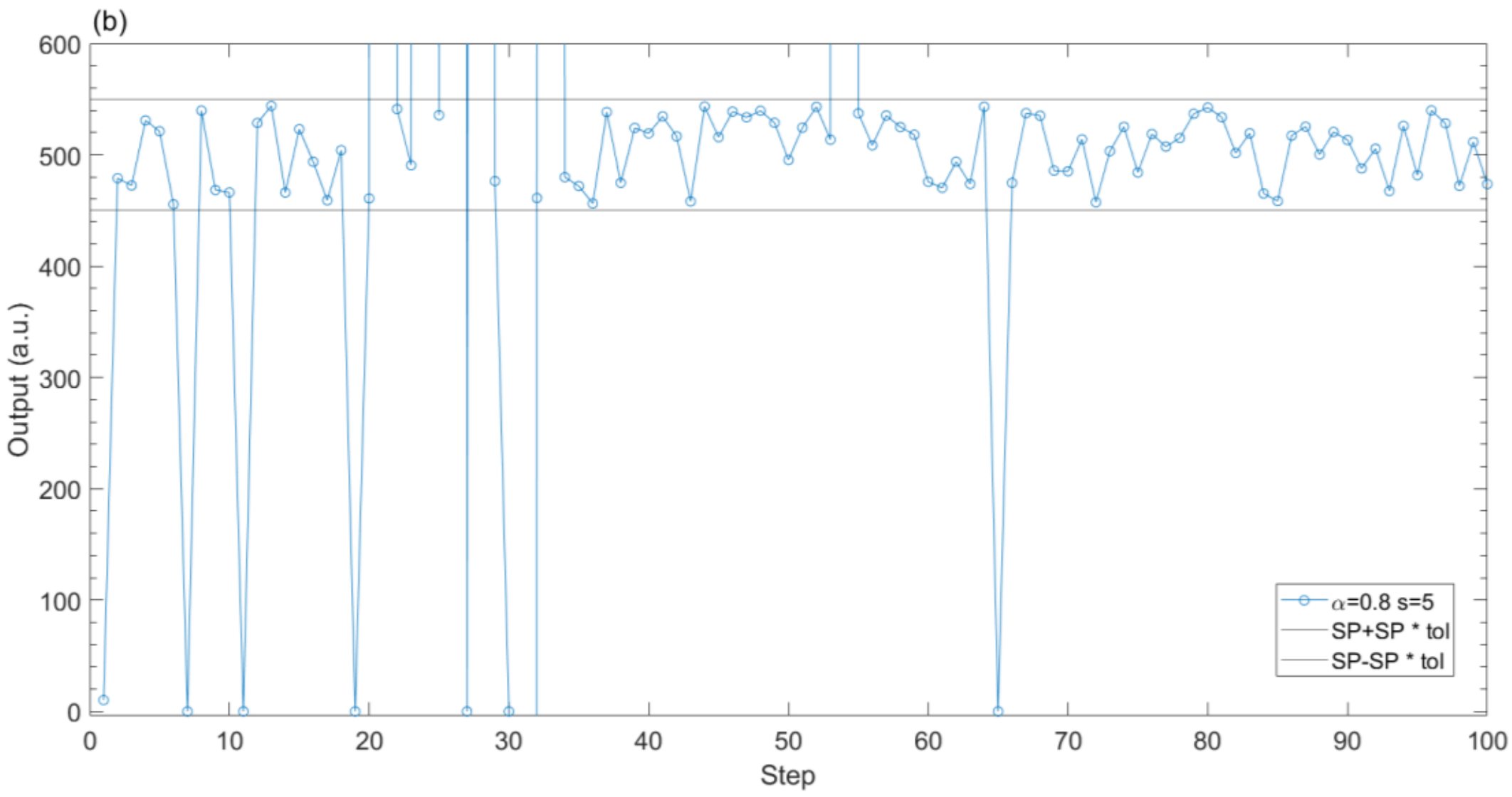

*Figure 5: System output evolution for 100 temporal steps for input evolution with $\alpha = 0.5$ and $s = 5$. Black lines represent the tolerance input around the target value. In (a) the evolution as result from update weight process without correlations; in (b) the evolution as result from update process with correlations. In the case without correlations the failures (output outside tolerance interval) are more frequent.*

For this last case in Figure 5the evolution of the system output under the weight update process is shown for one trial with 100 runs. In the case without correlations (panel a) the final result outside the acceptance interval around the set point are more frequent than the case with correlations (panel b).

For the sake of completeness, the results achieved with weights linearly dependent from inputs are shown in the following. The case with constant weights reduces to similar basic model (as the perceptron [22]) and it is not addressed here. The observed differences between weight update processes with and without correlations are still valid. Some differences are observed between linear and non-linear input dependence of weights. In Table 3 and Table 4 the same results obtained for the previous case are reported for the simulations run with the constraint of having linear dependence.

| $\alpha = 0.2$ | | |
|---|---|---|
| | $s = 1$ | $s = 5$ |
| With Correlation | 90.0 | 98.6 |
| Without Correlation | 84.5 | 74.2 |

| $\alpha = 0.8$ | | |
|---|---|---|
| | $s = 1$ | $s = 5$ |
| With Correlation | 83.7 | 95.0 |
| Without Correlation | 94.5 | 50.0 |

*Table 3: Table that present the average values of success rate for the case of linear input dependent weights.*

| $\alpha = 0.2$ | | |
|---|---|---|
| | $s = 1$ | $s = 5$ |
| With Correlation | 274 | 218 |
| Without Correlation | 629 | 873 |

| $\alpha = 0.8$ | | |
|---|---|---|
| | $s = 1$ | $s = 5$ |
| With Correlation | 726 | 497 |
| Without Correlation | 440 | 1334 |

*Table 4: Table that present the average number of realizations for the case of linear input dependent weights.*

As shown in the results, only under certain conditions of input parameters appreciable differences are observed. For a deeper insight the reader is reminded to the discussion section.

## 5. Discussion

The first modelling assumption in the application described above is the representation of weight functions as second degree polynomial. This class of functions was selected because they well capture several non-linear current-voltage characteristic curves in many classes of material and systems for neuromorphic applications and also some of non-linear electron conduction mechanisms [20], [21].

The main objective of this work is to characterize the proposed model and investigate its main features through a software-based simulation. The results show that tuning the degree of correlation among the weight updates has a direct influence on system performance. The performances are explored for different input conditions to avoid biases in the results. in particular different degree of correlations (tuning parameter $\alpha$) and different noise added as uniformly distributed number (tuning parameter $s$). It is evident that correlations exert a strong influence in case with $\alpha = 0.8$ and $s = 5$. In these conditions, correlations act as a stabilizing mechanism that prevents the system output from diverging and counterbalances the stochastic nature of the weight-update process. In some cases, where the degree of temporal correlation and noise scale in inputs are different, the runs without correlations show better results because in these cases the space of coefficient is sampled without constraints, giving the possibility to achieve the desired result faster.

All these results are linked to the polynomial structure of the weight functions. Although this is a particular implementation, the general character of the work is not lost. In fact, polynomials have a simple mathematical structure to be handled and properties of polynomial with random coefficients are well studied in the literature [23]. In addition, the weight update process based on coefficient sampling gives a clearer structure with polynomial functions, where coefficients enter linearly with respect to the independent variable. The stochastic sampling of the polynomial coefficients can be naturally formulated in terms of a covariance structure analogous to that used in Gaussian-process models [24]. On the other end the non-linearity introduced by quadratic terms introduce a curvature term on the input vector that allows to increase the space dimensionality of inputs in the same way of non-linear kernels [25]. This gives slightly different results in the system evolution for non-linear dependent weight (Table 1 and Table 2) from linear dependence (Table 3 and Table 4). In addition, these effects are combined with the different number of coefficients to be sampled for the two cases (the higher the number of coefficients to be randomly sampled the higher the possibility to give an output result far from the desired one). These two factors make plausible the differences observed in the two implementations. Anyway, in this part a lot of work has to be made to explore the analytical properties of the model; this will be object of a future publication.

## 6. Conclusions

In summary, a new model for input processing based on a non-conventional stochastic weighting process has been presented. The main part of the model is a non-linear input dependent weight matrix that map N input to M outputs and that can be updated though a random update process that does not require strong constraints training process. A software simulation of the model has been presented to highlight how the main features of the system, in particular the presence of correlation among different weights during the sampling process and non-linearity, can be tuned to achieve different performances.

The model has been built to capture some features of a new class of materials and systems that exploit multi-state and complex response to external stimuli (electrical, optical and chemical). The presented model has the double aim to present a new simple framework to develop new strategies from the side of algorithms and new design from the design of hardware. It is not based on a particular class of systems, but it is thought to be implemented by a heterogeneous class of materials.

The most urgent next steps of the work will be:

- The study of analytical properties of the model to understand approximation capabilities and a deeper understanding of the adaptation process
- The study of efficiency of an implementation of the model compared with state of art systems in the field of autonomous learning and system control for a class of selected tasks
- Understand how the model can represent new guidelines to design new hardware with relaxed constraint on fabrication/integration processes and more efficient implementation of neuromorphic systems.